\pdfoutput=1  % Fixes ERROR: no BoundingBox  from arXiv 
\documentclass[10pt, conference]{IEEEtran}
\IEEEoverridecommandlockouts
\usepackage{cite}
\usepackage{amsmath,amssymb,amsfonts}
\usepackage{graphicx}
\usepackage{textcomp}
\usepackage{xcolor}

\usepackage{rotating}

\usepackage[braket]{qcircuit}

\usepackage{lipsum} % Used for inserting dummy 'Lorem ipsum' text into the template

\usepackage[switch]{lineno}

\usepackage{listings} % for code text
\usepackage{tikz}

\usepackage{algorithm}
\usepackage{algpseudocode}

\newcommand{\zt}{$Z_2$}
\newcommand{\ztb}{\textbf{$\mathbf{Z}_2$}}

\begin{document}

\title{
Engineering quantum states with neutral atoms
}

\author{
\IEEEauthorblockN{Jan Balewski\IEEEauthorrefmark{1}, Milan Kornja\v{c}a\IEEEauthorrefmark{2}, Katherine Klymko\IEEEauthorrefmark{1}, Siva Darbha\IEEEauthorrefmark{1}\IEEEauthorrefmark{3}, Mark R. Hirsbrunner\IEEEauthorrefmark{1}\IEEEauthorrefmark{4},\\Pedro L. S. Lopes\IEEEauthorrefmark{2}, Fangli Liu\IEEEauthorrefmark{2}, Daan Camps\IEEEauthorrefmark{1}}
\IEEEauthorblockA{\IEEEauthorrefmark{1}National Energy Research Scientific Computing Center, Lawrence Berkeley National Laboratory, Berkeley, CA, USA}
\IEEEauthorblockA{\IEEEauthorrefmark{2}QuEra Computing Inc., Boston, MA, USA}
\IEEEauthorblockA{\IEEEauthorrefmark{3}Applied Mathematics and Computational Research Division, Lawrence Berkeley National  Laboratory, Berkeley, CA , USA}
\IEEEauthorblockA{\IEEEauthorrefmark{4}Dept. of Physics and Institute for Condensed Matter Theory, University of Illinois at Urbana-Champaign, Urbana, IL, USA}
\IEEEauthorblockA{Email: balewski@lbl.gov}
}

% \author[1,*]{Jan Balewski\thanks{*Corresponding author: balewski@lbl.gov}} %  balewski@lbl.gov, ORCID: 0000-0002-1899-6526
% \author[4] {Milan Kornja\v{c}a} % mkornjaca@quera.com
% \author[1]{Katherine Klymko } % kklymko@lbl.gov , ORCID  0000-0002-4158-5776
% \author[2,1]{Siva Darbha} % sdarbha@lbl.gov
% \author[3,1]{\\Mark R. Hirsbrunner} % MarkHirsbrunner@lbl.gov, mark@mhirsbrunner.com
% \author[4] {Pedro Lopes}  % plopes@quera.com
% \author[4] {Fangli Liu} % fliu@quera.com
% \author[1]{Daan Camps } % dcamps@lbl.gov, ORCID 0000-0003-0236-4353

% \affil[1]{National Energy Research Scientific Computing Center,
% Lawrence Berkeley National Laboratory, Berkeley, CA, USA}
% \affil[2]{Applied Mathematics and Computational Research Division, Lawrence Berkeley National  Laboratory, Berkeley, CA , USA}
% \affil[3]{Department of Physics and Institute of Condensed Matter Theory, University of Illinois at Urbana-Champaign, Urbana, IL, USA}
% \affil[4]{QuEra Computing Inc., Boston, MA, USA}

\maketitle
%\linenumbers  % activate here

\begin{abstract}
 Aquila, an analog quantum simulation platform developed by QuEra Computing, supports control of the position and coherent evolution of up to 256 neutral atoms. This study details novel experimental protocols designed for analog quantum simulators that generate Bell state entanglement far away from  the blockade regime, construct a \ztb\ state with a defect induced by an ancilla, and optimize the driving fields schedule to prepare excited states with enhanced fidelity.
 We additionally evaluate the effectiveness of readout error mitigation techniques in improving the fidelity of measurement results. All experiments were executed on Aquila from QuEra and facilitated by the AWS Braket interface. 
 Our experimental results closely align with theoretical predictions and numerical simulations. The insights gained from this study showcase Aquila's capabilities in handling complex quantum simulations and computations, and also pave the way for new avenues of research in quantum information processing and physics that employ programmable analog hardware platforms.
\end{abstract}


%%%%%%%%%%%%%%%%%%%%%%%%%%%%%%%%%%%%%%%%%%%%%%%%%%%%%%
%%%%%%%%%%%%%%%%%%%%%%%%%%%%%%%%%%%%%%%%%%%%%%%%%%%%%%

% !TEX root = 0-main.tex
\section{Introduction}

Analog quantum computing systems based on neutral atoms offer a promising platform for exploring quantum many-body dynamics~\cite{Browaeys:2020}. These systems allow investigations of one- and two-dimensional phenomena~\cite{Bluvstein:2021} and led to the discovery of quantum many-body scars~\cite{Bernien:2017,Turner:2018}. They have facilitated experimental studies into the quantum Kibble-Zurek mechanism for second-order phase transitions~\cite{Keesling:2019,Ebadi2021}. Furthermore, such systems are instrumental in probing ground state phase diagrams~\cite{Chen:2023} and examining topological phases, including topological spin liquids~\cite{Semeghini:2021, Kornjaca:2023}. Additionally, they offer a natural framework for experimentally simulating lattice gauge theories~\cite{Surace:2021} and solving the maximum independent set (MIS) problem on unit disk graphs~\cite{Ebadi:2022,Nguyen:2023, finzgar:2023}. Lastly, they have shown potential for quantum error correction ~\cite{Bluvstein2022,Bluvstein2024}, making use of the programmable and scalable controls. Recently, these systems have become publicly available through a cloud access model~\cite{quera_amazon_braket}. 

The aim of this study is to perform analog quantum computation on a neutral atom system for several proof-of-concept quantum simulation and optimization tasks. Throughout our experiments, we test novel ancilla-assisted geometries, parallel programming capabilities, and optimal control techniques that enable higher fidelity quantum simulations. In particular, we present innovative experimental approaches to generate entanglement induced by interaction beyond the blockade regime. Our work then proceeds to construct a \zt\ state with a defect using stray fields from an ancilla. Finally, we fine-tune the driving field schedule to improve the preparation of states not corresponding to the lowest energy state, i.e. the Maximal Independent Set (MaxIS) problem. All of our experiments use the Aquila system~\cite{wurtz2023aquila} accessed through the AWS Braket interface~\cite{quera_amazon_braket}.
% !TEX root = 0-main.tex
\section{The Aquila Neutral Atom Quantum System}

The Aquila quantum processor~\cite{wurtz2023aquila}
uses an array of $^{87}$Rb atoms,  which are trapped using optical tweezers, to serve as qubits.
The processor allows for the precise positioning and coherent evolution of up to 256 atoms in a nearly arbitrary, user programmable, two-dimensional configuration. Furthermore, users can program the quantum simulation protocol through the Rabi frequency, phase, and detuning frequency of the driving laser.

As such, Aquila can perform coherent time-evolution under the {\it Rydberg Hamiltonian}~\cite{wurtz2023aquila},
\begin{align}
    \frac{\mathcal{H}(t)}{\hbar} &= 
        \frac{\Omega(t)}{2} \sum_j \left( e^{i \phi(t)} \ket{g_j} \bra{r_j} + e^{-i \phi(t)}  \ket{r_j} \bra{g_j} \right) \nonumber \\
      &\quad ~~~- \Delta(t) \sum_j \hat{n}_j 
       + C_6 \sum_{j < k}\frac{\hat{n}_j \hat{n}_k}{x_{jk}^6},
\label{eq:rydb-hamilt}
\end{align}
with $\Omega(t)$ and $\Delta(t)$ the Rabi frequency and detuning of the driving laser, respectively. The phase $\phi(t)$ is set to zero throughout all demonstrations in this work. The laser field couples the ground state $\ket{g_j}$ of the $j$th atom to the Rydberg state $\ket{r_j}$. A Rydberg state is a highly-excited state of an atom, characterized by a high principle number and large spatial extent. The final term in Eq.~\eqref{eq:rydb-hamilt} is the two-body van der Waals interaction, which decreases rapidly with the distance $x_{jk} = \lVert \vec x_j - \vec x_k \rVert$ between two atoms $j$ and $k$. The number operator $\hat{n}_j = \ket{r_j}\bra{r_j}$ for the $j$-th atom is 1 if it is excited to the Rydberg state and 0 if it is in the ground state. The value of the constant $C_6$ is  $862,690 \times 2\pi $~MHz~$\mu \text{m}^6$ for the choice of $^{87}$Rb atoms.

The distance at which the van der Waals force exceeds the Rabi drive 
is known as the blockade radius $R_b$, and is given by,
\begin{align}
R_b&= \left(\frac{C_6}{ \sqrt{\Omega^2 + \Delta^2}} \right)^{1/6}.
\label{eq:Rb}
\end{align}
When two atoms are within a distance $R_b$ of each other, simultaneous excitation of both atoms into the Rydberg state is strongly penalized. The Rydberg blockade phenomenon enables encoding conditional logic through the means of the geometry of the atom array.

In the majority of the quantum protocols discussed in this paper, we begin by rapidly increasing the Rabi frequency $\Omega(t)$ from zero to its experimental maximum and afterwards linearly  ramping the detuning $\Delta(t)$ while keeping the Rabi frequency constant. By varying the detuning ramp duration, $t_R$, we control the quantum evolution. Additionally, for some experiments, we also program a non-linear schedule of $\Delta(t)$ to further fine-tune the dynamics. In all experiments we report results as a function of the total evolution time $T$, which includes the detuning ramp duration and the additional time required to increase and decrease the Rabi frequency.

% !TEX root = 0-main.tex
\section{Single-atom Rabi Oscillations}

Rabi oscillations describe the coherent evolution of a quantum system, such as an atom or a qubit, between two energy states when subjected to a resonant electromagnetic field. This oscillation manifests the quantum superposition principle.
The ability to demonstrate Rabi oscillations can be viewed as a ``Hello world'' program for a quantum processor, serving as a good check of the single-qubit evolution fidelity, the qubit relaxation time, and the readout errors.

\begin{figure}[htbp]
\centerline{\includegraphics[width=.99\linewidth]{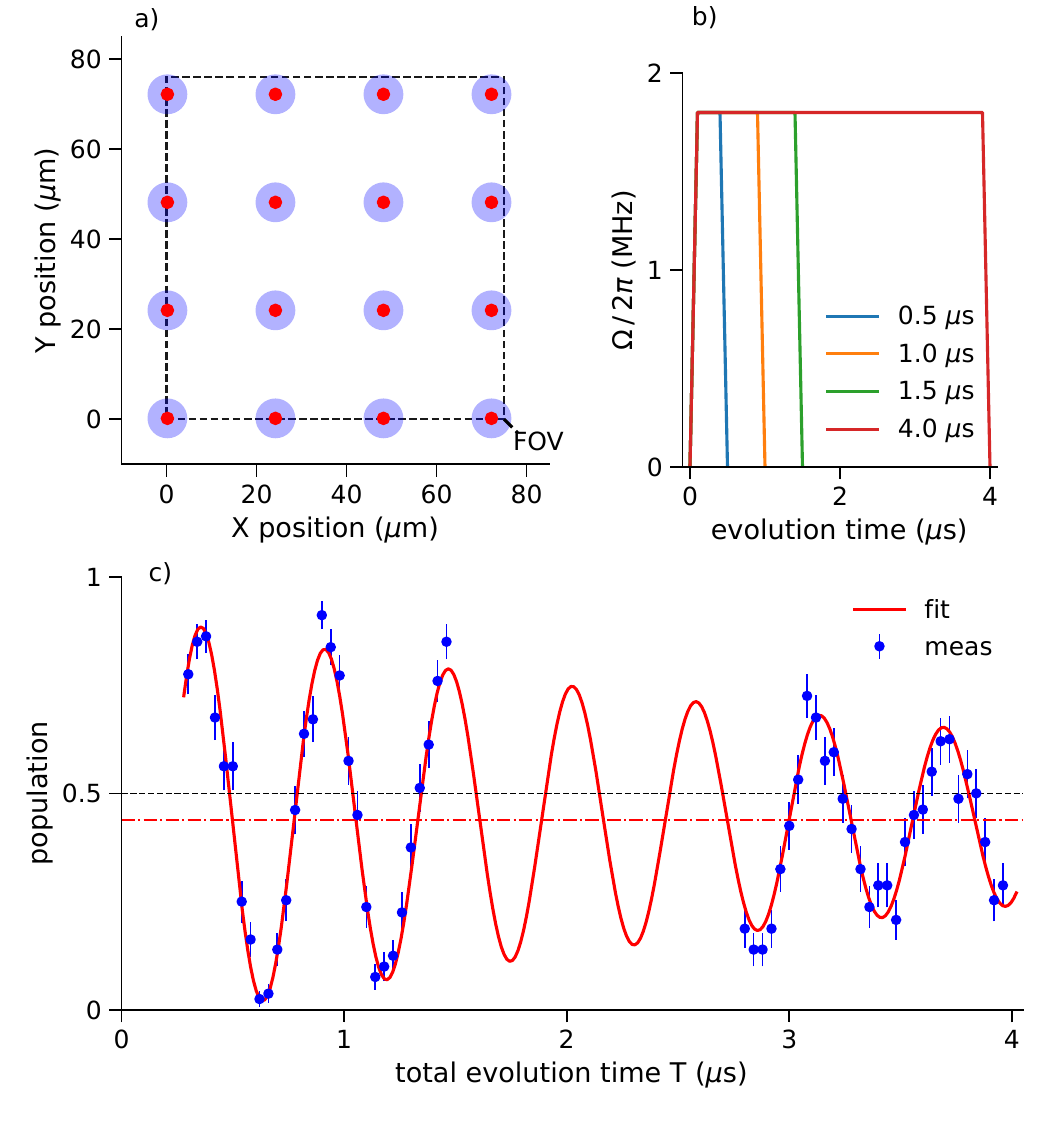}}
   \caption{ Rabi oscillation experiments. a) The placement of 16 atoms  within the field of view (FOV) of Aquila. The blue circles have diameter equal to the Rydberg blockade radius. b) Examples of the $\Omega(t)$ schedule for selected values of the total evolution duration, $T$. c) Measured data (blue circles, with error bars) and fitted function Eq.~\eqref{eq:sin_exp} (solid red curve). The red dashed-dotted horizontal line represents the fitted parameter $C$. The time value $T$ denotes the evolution time in panel b), when the $\Omega$ drives return to zero.
}
\label{fig:rabi}
\end{figure}

We demonstrate Rabi oscillations using Aquila configured with 16 neutral atoms arranged on a square lattice with an interatomic distance much greater than the blockade radius, as illustrated in Fig.~\ref{fig:rabi}a. The driving field schedule, $\Omega(t)$, depicted in Fig.~\ref{fig:rabi}b, includes a 0.1~$\mu$s ramp-up, a variable-duration  plateau at a strength of $\Omega_\mathrm{max}=1.8 \times 2\pi ~$MHz, and a 0.1~$\mu$s ramp-down. 
The detuning $\Delta(t)$ is set to zero throughout the evolution. The choice of $\Omega_\mathrm{max}$ corresponds to a Rydberg blockade radius of $R_b=8.8~\mu$m, visualized in Fig.~\ref{fig:rabi}a as the blue circles  with radii equal to $R_b/2$. We position the atoms $d=24~\mu$m apart, such that the Rydberg interaction between them becomes negligible and they evolve independently from each other. We measure the Rabi oscillations across 16 atoms simultaneously to test the parallel programming capabilities of the hardware.

We perform 60 experiments sampling the total duration $T$ of the $\Omega$ drive within two intervals: $[0.3,1.5]~\mu$s and $[2.8,4.0]~\mu$s, which covers seven Rabi oscillation cycles~\footnote{ The $T$ values below $0.3\mu$s are experimentally inaccessible because of the mandatory ramp-up and ramp-down periods.}. Each configuration is measured 5 times (thus collecting 5 ``shots"), and the probability of observing the Rydberg state is averaged over the 16 atoms. Because the atoms evolve independently, this is equivalent to 80 shots for each value of $T$.  The measured population of the Rydberg state is depicted as a function of $T$ by the data points in Fig.~\ref{fig:rabi}c. 

The data clearly shows the Rabi oscillations, albeit with reduced amplitude, attributable to Aquila's finite coherence time and  readout errors. A damped sinusoidal function,
\begin{align}
f(t) &= C + A \sin(\omega t + \phi) \cdot \exp(-t/\tau),
\label{eq:sin_exp}
\end{align}
accurately fits the observed data, as indicated by the red curve, with a resulting $\chi^2/\textrm{DOF}$  ratio~\footnote{$~\chi^2$ measures the error-weighted discrepancy between observed data and a parameterized model function, while $\chi^2/\textrm{DOF}$, with DOF indicating degrees of freedom, normalizes this discrepancy. A good fit is characterized by $\chi^2/\textrm{DOF} \approx 1$.}  of 1.2. The measured relaxation time was $\tau=4.5 \pm 0.5 ~\mu$s. 
This result is modestly better than the value $\tau \approx 3.6 ~\mu$s reported in~\cite{wurtz2023aquila}.

%==========================================
% JAN, DO NOT ERASE IT : ~/paper_QCE2024_engineering$ ./fig1_rabiOscil.py
 %========================================

% !TEX root = 0-main.tex
\section{maximally entangled two-qubit state}

%==========================================
% JAN, DO NOT ERASE IT : ~/paper_QCE2024_engineering$ ./fig2_bellState.py
\begin{figure*}[htbp]
\centerline{\includegraphics[width=.99\linewidth]{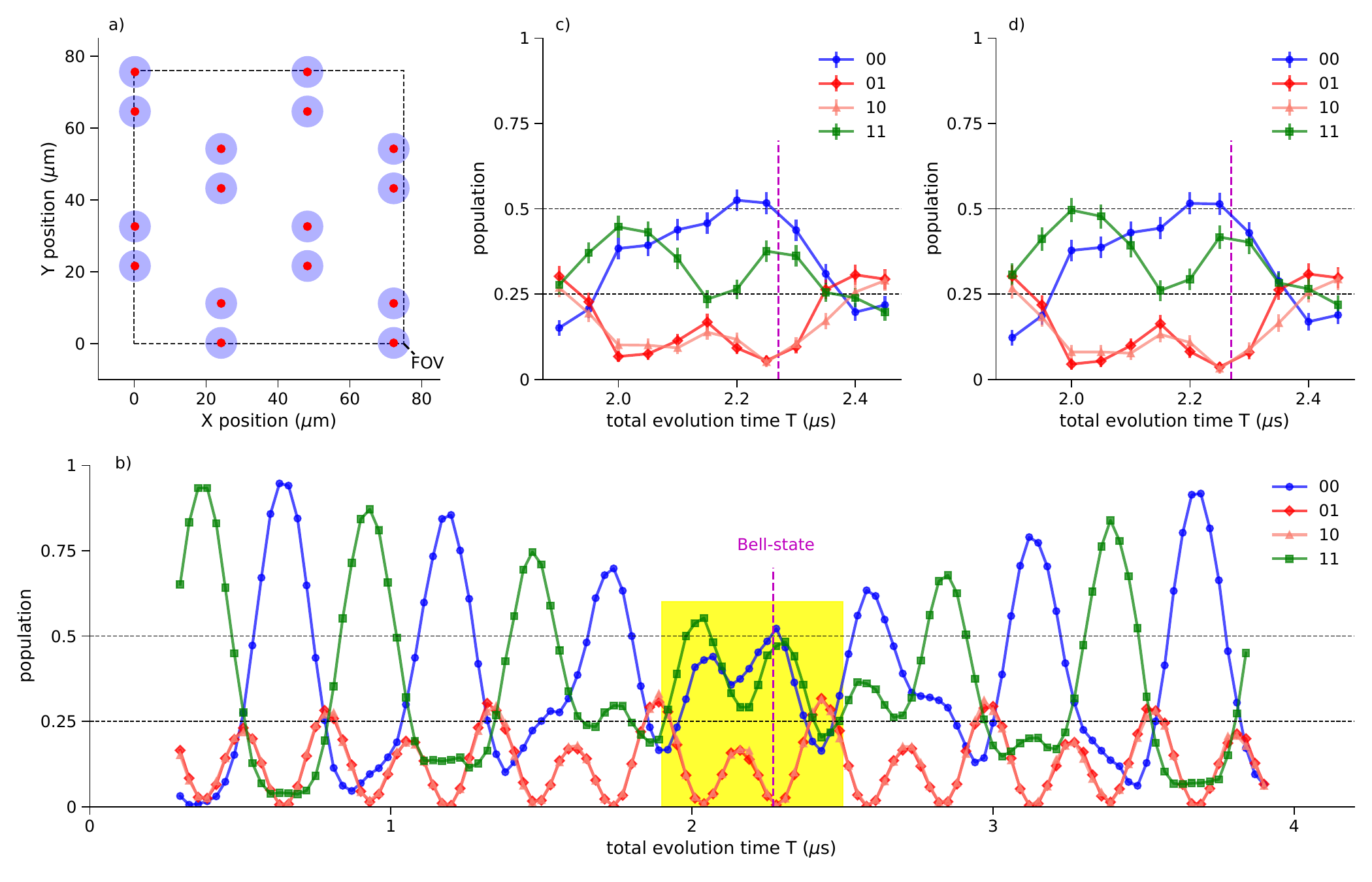}}
   \caption{ Bell-state preparation experiment. a) The placement of atom pairs in the FOV of Aquila, see also caption of Fig.~\ref{fig:rabi}a. b) Simulation of the population of 4 measured bitstrings vs. total evolution time. The vertical magenta line marks the location of an approximate  Bell state. c)  Raw experimental results from Aquila. d) Experimental results after readout error mitigation. 
}
\label{fig:bell}
\end{figure*}
 %========================================

In the realm of quantum information processing, generating entangled states is fundamental to proving that we are operating on a quantum system. The paradigmatic examples of entangled states are the two-qubit Bell states, which are an equal superposition of $ \ket{00}$ and $\ket{11} $ states.
Previous research~\cite{PhysRevLett.123.170503} showed how to implement a controlled-phase entangling gate using neutral atoms, essential for constructing  Bell states, by employing a fast protocol that relies on global coupling of two closely positioned atoms, briefly excited to Rydberg states. In this study, we adopt a slower protocol to achieve a two-qubit entangling gate, operating in the weak interaction regime with atoms positioned outside the Rydberg blockade area.

In our Bell state experiment, we consider pairs of atoms with an interatomic distance of $11~\mu$m and $R_b=8.6~\mu$m. At this distance, the Van der Waals interactions, described by the last term in Eq.~\eqref{eq:rydb-hamilt}, facilitates weak coupling between the atoms. The Rabi frequency and the resulting blockade radius are the same as in the previous section. We use a configuration of eight atom pairs that are spaced sufficiently far apart to evolve independently, as depicted in Fig.~\ref{fig:bell}a. As a result, we execute eight concurrent entangling gates. For multi-atom measurements, we adopt the bitstring notation, which assigns a `1' to an atom measured in the excited state $\ket{1}$ and a `0' to the ground state $\ket{0}$. The order of the bits corresponds to the indexing of the atoms.

To predict the evolution of an atom pair, we conducted classical simulations using the Braket SDK~\cite{quera_amazon_braket}. Atoms were initialized in the ground state. We incremented the duration of the Rabi drive, $\Omega$, in steps of 30~ns and computed the probability of the four possible bitstrings. Fig.~\ref{fig:bell}b presents the simulation outcomes, highlighting that at $T=2.3 ~\mu$s the probabilities for the `00' and `11' states are each approximately 0.5, while the populations in the `01' and `10' states are near zero, indicating the creation of a Bell state.

The physical principle underlying our preparation of the Bell state relies on  the subtle difference in the periodicity of the probabilities for the `00' and `11' populations. This difference gradually leads to phase accumulation at a rate  determined by the Rydberg interaction.
The timescale for performing Bell state preparation can be estimated in the small interaction limit, $V_r= C_6/d^6 \ll \Omega$. In the non-interacting limit, the state $\ket{v}=\left(\ket{00}-\ket{11}\right)/\sqrt{2}$ is an eigenstate with energy zero. Introducing the Rydberg interaction perturbatively gives the first order energy correction of $V_r/2 \approx 0.24~$MHz which, up to next order corrections in $V_r/\Omega$, is the frequency of the beating between `00' and `11' that we observe. In order to obtain the Bell state, a beating phase of $\pi$ has to be accumulated, giving us the first order estimate of $t \approx 1/V_r \approx 2.08 \mu$s, very close to the observed preparation time that can be obtained from the full calculation.

% Based on the simulation and theoretical insights on the effective duration of the Rabi drive $\Omega$ for achieving the Bell state, 
Based on the simulations and theoretical insights above, 
we conducted measurements at 12 total drive times covering the range $T \in [1.9,2.5]~ \mu$s, corresponding to the highlighted region in Fig.~\ref{fig:bell}b. The experimentally observed probabilities, as shown in Fig.~\ref{fig:bell}c, qualitatively agree with the simulations. At $T=2.3 ~\mu$s, the  measured `01' and `10' state probabilities were notably lower than predicted. However, the probability for `00' was observed to be 25\% higher than that for `11'. We attribute this difference to the readout errors of Aquila that are biased towards mistakenly detecting the `1' state as `0', \cite{wurtz2023aquila}.

To mitigate the readout error effects, we applied 
a variant of the tensor product error mitigation procedure described in~\cite{Bravyi:2021}. 
Our experimental insights into the hardware led us to adopt a specific error model. Firstly, we consider a 5\% probability (\(\epsilon=0.05\)) that an atom in state `1' might be mistakenly read as being in state `0'. Conversely, we assume that an atom in state `0' is always measured accurately as being in state `0', \cite{wurtz2023aquila}. The asymmetry in the readout errors stems from the exclusive decay of the highly excited Rydberg state during measurement. \cite{Bernien:2017} Additionally, we treat these errors as statistically independent, which is a reasonable assumption for isolated atoms being detected through fluorescence.  Lastly, we anticipate that each set of measurement outcomes (bitstrings) might have at most one error in the bit value, ignoring the \(\mathcal{O}(\epsilon^2)\) effects.

The results obtained after error mitigation are illustrated in Fig.~\ref{fig:bell}d and show a clear increase in the population of the `11' state in the vicinity of $T=2.3~\mu$s. While we can't readily measure the Bell state phase to estimate preparation fidelity using the current Hamiltonian-based Aquila control setup that allows for only computational basis measurements, the mitigation clearly improved the population ratio of `11' and `00' bitstrings, bringing them closer to the expected values.

This experiment not only demonstrated the capability of Aquila  to prepare a Bell state but also verified the spatial uniformity across Aquila's field of view (FOV), as we conducted the Bell state experiment concurrently on 8 qubit pairs distributed throughout the entire FOV. If  variation across the FOV was significant, it would necessitate a much larger $\epsilon$, in order to reduce the probabilities of both the  `01' and `01' states to approximately null. 

% !TEX root = 0-main.tex
\section{ Engineering the \zt\ state with defects via an optimized drive}

In quantum many-body physics, the \zt\  phases and phase transitions represent strongly correlated phenomena that, due to their simplicity and universality, have formed the basis of our understanding of quantum phase transitions~\cite{Sachdev:2011}. Neutral atom systems have already been extensively used to study these fundamental phases by adiabatically preparing 1D~\cite{Bernien:2017,wurtz2023aquila} and 2D~\cite{Ebadi2021} antiferromagnetic \zt\ states. Here, we optimize protocols to prepare a \zt\ phase with a domain-wall defect in a chain of atoms using an ancilla atom.

\begin{figure}[htbp]
\centerline{\includegraphics[width=.99\linewidth]{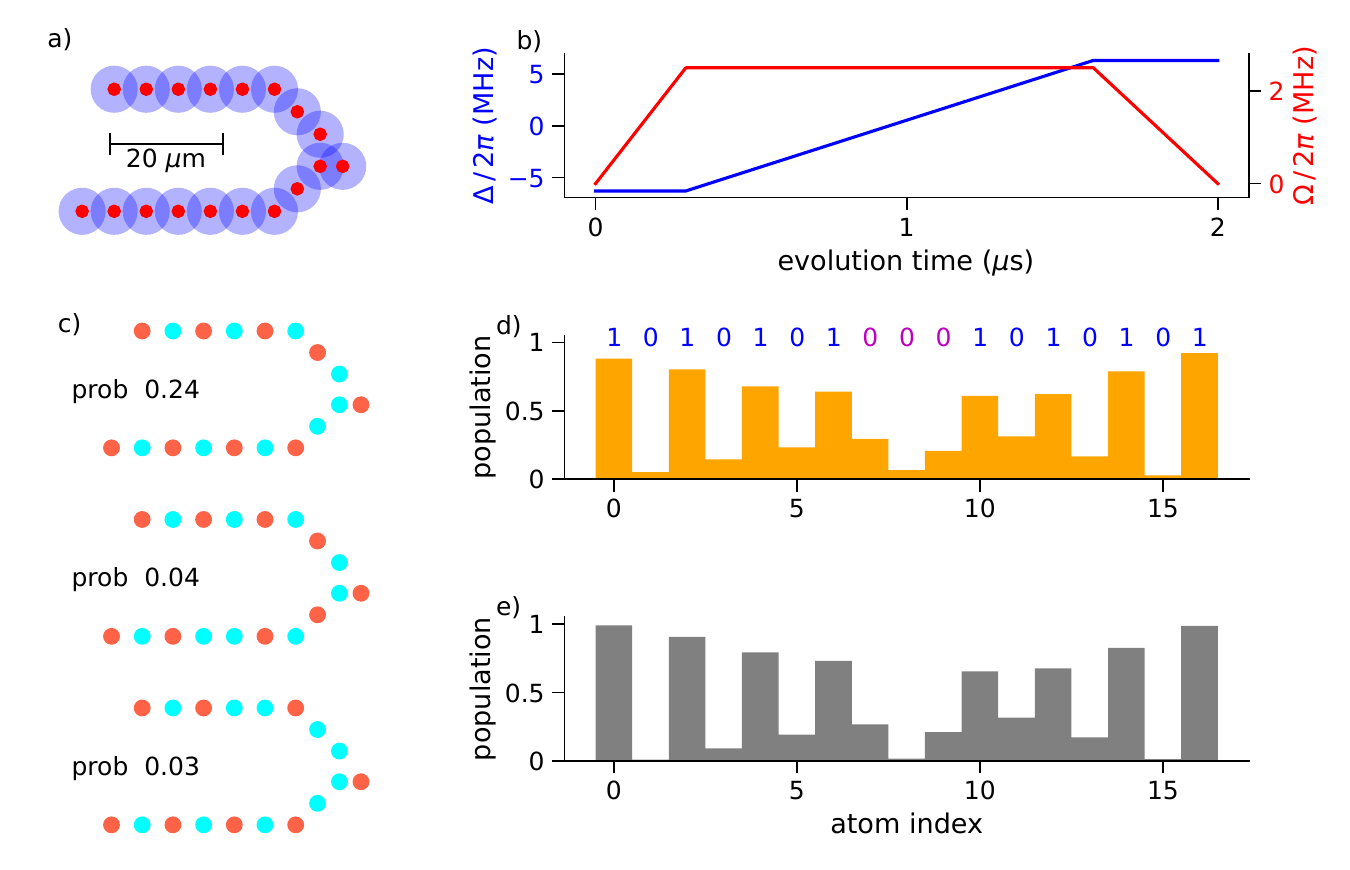}}
   \caption{\zt\ experiment without drive optimization. 
   a) The 1D atom configuration. The amount of overlap between the blue circles illustrates the strength of the Rydberg interaction if both neighbouring atoms are in the excited state. b) The linear detuning drive is shown in blue, while the Rabi drive schedule is in red. c) The most probable experimentally observed patterns, where red (blue) means the 1 (0) state. d) Experimental probability of the 1-state vs. atom index in the chain. The target classical pattern is printed above the histogram.  e) The simulated probability of the 1-state for the linear drive in panel b.
}
\label{fig:uturn17a lin}
\end{figure}

For the \zt\  experiment we have chosen a 17-atom chain, shown in Fig~\ref{fig:uturn17a lin}a.   The chain 
forms a U-turn shape to fit in the Aquila FOV. The 18-th ancilla atom on the farthest right will bias the central atom in the chain to stay in the ground state.
We simultaneously apply a global detuning field, \(\Delta(t)\), and a Rabi field, \(\Omega(t)\), illustrated in Fig.~\ref{fig:uturn17a lin}b by blue and red lines, respectively. The purpose of these two fields is to induce an adiabatic transition from the initial state, where all atoms are in the ground state, to a target state. The \(\Omega\) field induces mixing of atoms between the ground and Rydberg states. %, providing the necessary energy for this transition.
The Rydberg interactions, primarily occurring in pairs, promote a specific pattern of alternating states along the chain. Additionally, the detuning field is applied to maximize the fraction of excited atoms in the system. At the evolution's conclusion, the \(\Omega\) field is deactivated, effectively `freezing' the atoms in their final states. The experiment ends with a simultaneous measurement of the states of all atoms involved.

This experiment is repeated 1000 times, resulting in a categorization of observed bit patterns. Fig.~\ref{fig:uturn17a lin}c shows the three most probable final states. Notably, the target pattern featuring `000' in the center has a probability of 0.24. This pattern has the maximum number of 9 Rydberg excitations, marked in red.
The next two most probable states have higher energy  and are recorded  six times less frequently. Given that the random chance for any pattern is approximately \(2^{-17}\sim 10^{-5}\), these results underscore the capability of the neutral atom Quantum Processing Unit to select the minimum energy state and several lower excited states.% for an exponentially complex combinatorial problem. 

The aggregated population distribution across all measurements is presented in Fig.~\ref{fig:uturn17a lin}d, highlighting a reduced probability for the three central atoms in the sequence. Meanwhile, the characteristic \zt\  alternating pattern is visible across all other positions.

Fig.~\ref{fig:uturn17a lin}e, which displays the classically simulated population distribution, closely resembles the experimental results from panel d. This similarity confirms that the model Hamiltonian of Aquila, used in the classical simulation, aligns well with the actual hardware Hamiltonian.%, indicating that the unaccounted-for experimental noise is minimal.

%==========================================
% JAN, DO NOT ERASE IT : ~/paper_QCE2024_engineering$  ./fig3_driveOpt.py -t lin
\begin{figure}[htbp]
\centerline{\includegraphics[width=.99\linewidth]{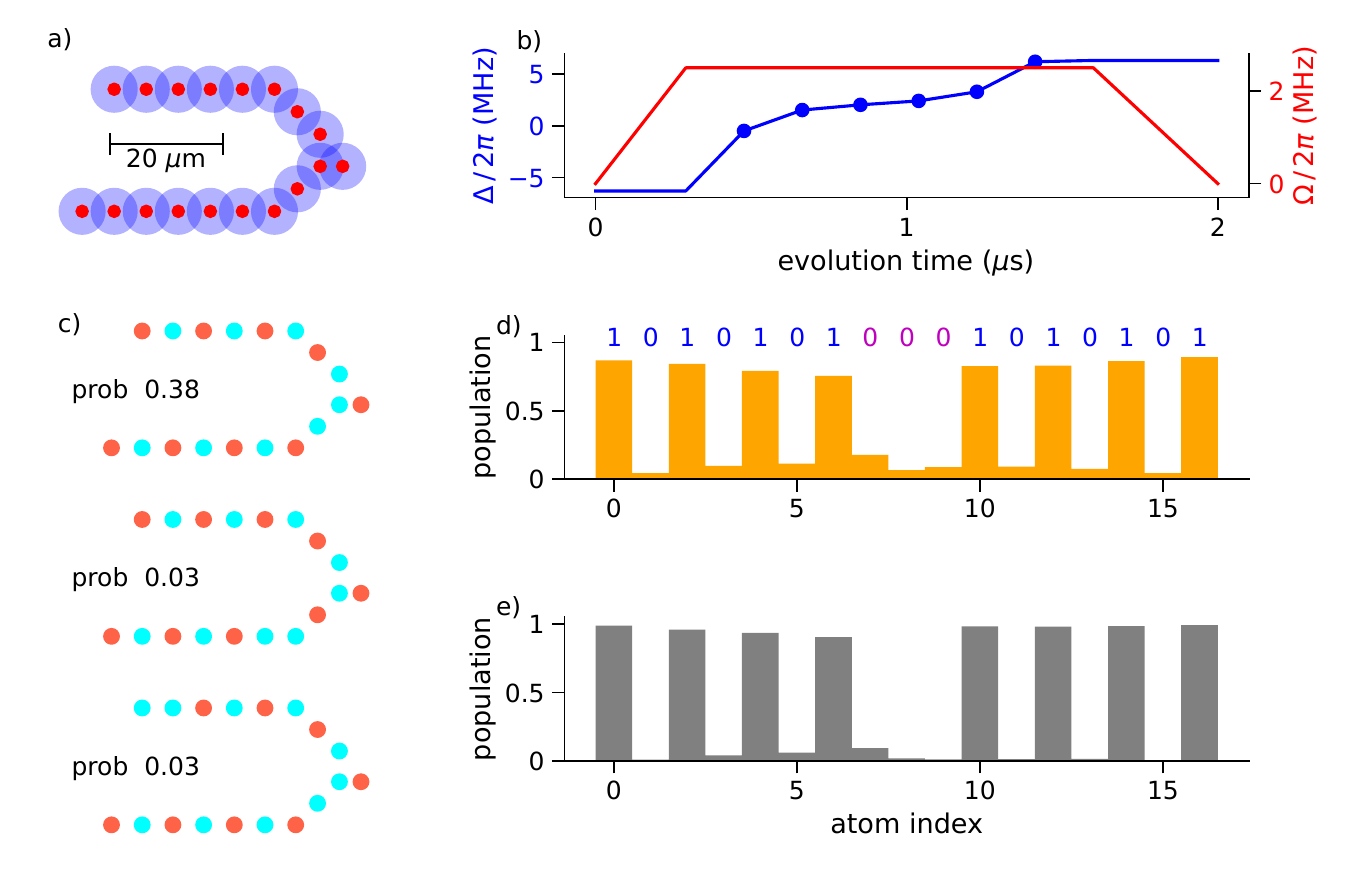}}
   \caption{\zt\ experiment with drive optimization.  a) Atom configuration. b) The optimized  detuning drive. c) The most probable experimentally observed patterns. d) and e) Probability histogram of 1-state vs. atom index for experiment and simulations, respectively.
}
\label{fig:uturn17a opt}
\end{figure}

 %========================================
Can we achieve a higher probability of reaching the target state with the `000' pattern in the middle of the chain? Yes, by optimizing the detuning drive schedule. Due to technical limitations, optimization is conducted using the Braket classical  simulator for a 17-atom chain plus one ancilla. The COBYLA optimizer~\cite{Powell1994COBYLA} is selected for its ability to operate without optimizing the gradient of the function, a challenging computation in shot-based settings. The resulting optimized detuning schedule is depicted in Fig.~\ref{fig:uturn17a opt}b in blue. We adopt a step-wise linear parameterization for the detuning, allowing for arbitrary amplitude changes at six uniformly spaced and fixed time intervals, marked by blue circles. The durations of the ramp-up and ramp-down intervals are set at 0.29 and 0.4~$\mu s$, respectively. For each detuning schedule instance, 200,000 simulations (a.k.a shots) are performed, and the probabilities of all observed patterns are cataloged. COBYLA successfully maximizes the probability of the target 17-bit pattern featuring `000' in the center: \( '1010101{\bf 000}1010101' \). Approximately 50 iterations are necessary for COBYLA to converge.

The shape of the optimized drive is consistent with general expectations from the adiabatic theorem, by which the number of defects in the state depends most strongly on the speed of the passage through the minimum gap point of the instantaneous Hamiltonian spectrum. The position of the minimum gap point for the infinite chain is around $\Delta/\Omega \approx 1$ \cite{Keesling:2019}; while our finite configuration with a side atom is significantly different, the slowdown observed in the optimized protocol is qualitatively consistent with this estimate, with some shift to smaller  $\Delta/\Omega$ likely due to the presence of the domain-wall defect requiring lower overall Rydberg density in the system.

Utilizing this optimized detuning drive, a 2nd experiment featuring the same  17-atom chain, see Fig.~\ref{fig:uturn17a opt}a, is conducted. This new approach increases the probability of obtaining the target pattern to 0.38, a 1.5-fold improvement over the unoptimized linear detuning. Fig.~\ref{fig:uturn17a opt}c displays the three most probable patterns, with the probabilities of the second and third patterns remaining at 0.03. Consequently, this approach doubles the relative probability of achieving the target solution compared to alternatives, highlighting that a more significant probability gap is achievable with an optimized drive. This enhancement in isolating the target state is further evidenced by the aggregated population distribution, shown in Fig.~\ref{fig:uturn17a opt}d, where a markedly lower average probability in the middle of the chain and a more uniform alternating pattern across other atoms are observed, relative to Fig.~\ref{fig:uturn17a lin}d. We again note the classical simulations, shown in panel e are in good agreement with the experiment shown in panel d.

The success of the protocol for preparing the defect-full \zt\ state is underpinned by the two programmable controls in Aquila. On one hand, the programmable connectivity allows us to control excitations in the main chain, while on the other the programmable pulse protocols allow for the implementation of the optimized adiabatic drive. Our result opens up the possibility to further study the quantum dynamics of the prepared pinned defects in neutral atom hardware.

We note that this optimization was performed numerically on a smaller-sized system than the ones used in experiment. However, it appears to transfer quantitatively to the hardware results because the energy gaps did not change significantly when two atoms were added.

%==========================================
% JAN, DO NOT ERASE IT : ~/paper_QCE2024_engineering$  ./fig3_driveOpt.py -t opt
 %========================================

% !TEX root = 0-main.tex
\section{Finding  maximal independent sets}

An independent set (IS) is a set of vertices in a graph for which no two vertices within the set are adjacent.
The maximal IS (MaxIS) additionally requires  that adding any additional vertex to the set would violate its independence.  
The efficient search for a MaxIS
has applications across various domains. 
For example, a MaxIS can be utilized to identify non-overlapping resources that can be allocated to different entities without conflict, thereby optimizing utilization and avoiding interference. This approach is beneficial in diverse scenarios, such as preventing interference between nodes transmitting data packets~\cite{Purohit2014ConstructingMA,Afek2013}, and in graph coloring algorithms~\cite{wurtz2024industry}.

Neutral atom quantum processors are adept at finding the IS of maximum  cardinality, known as the Maximum Independent Set (MIS) problem \cite{Ebadi:2022, Nguyen:2023, finzgar:2023}, which maps to the unit-disk-graphs. The MIS solution proceeds by mapping the nodes directly to the atoms and imposing constraints through the Rydberg blockade that only allow IS solutions. If the detuning is large, an MIS solution is the ground state of the Rydberg Hamiltonian with zero Rabi frequency, and an adiabatic protocol can be attempted to prepare it. While a direct mapping of the general MaxIS problem to the ground state of the Rydberg Hamiltonian does not exist natively, we have investigated the feasibility of employing non-adiabatic protocols to solve the MaxIS problem using a neutral atom processor.

\begin{figure}[htbp]
\centerline{\includegraphics[width=.99\linewidth]{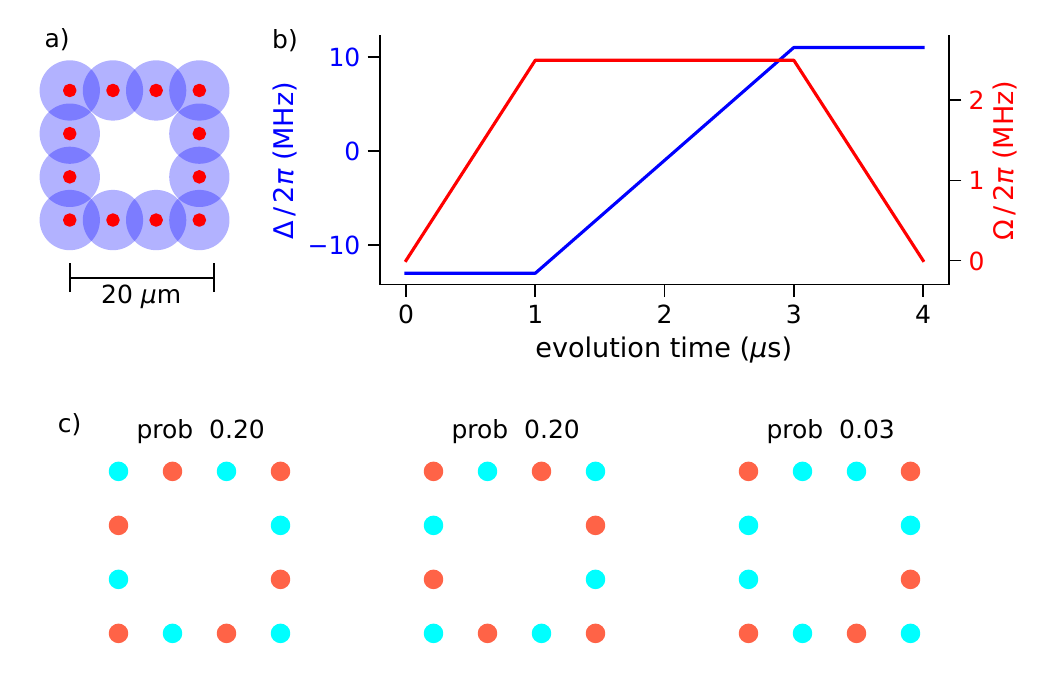}}
   \caption{MIS experiment without drive optimization. a) The placement of 12 strongly interacting atoms. The atoms represent the nodes in the corresponding graph problem and overlapping blockade radii the graph edges. b) The adiabatic drive used to search for an MIS solution. c) The 3 most probable solutions found by Aquila. The orange atoms are excited and represented the MIS solution.  The first 2 are correct MIS solutions for this problem. 
}
\label{fig:MaxLIS lin}
\end{figure}

We developed such a protocol and applied it on Aquila to a small MaxIS problem -- a 12-node loop-graph depicted in Fig.~\ref{fig:MaxLIS lin}a. The loop-connectivity is mapped to the Rydberg Hamiltonian, as can be seen from the overlapping blockade radii shown in blue.

 It is apparent upon analysis that a twelve-node loop-graph contains two MIS solutions of size six, both of which exhibit the same minimum energy of the Rydberg Hamiltonian. We first validate the ability of Aquila to find both MIS solutions by applying a linear and slowly varying detuning, facilitating the adiabatic evolution, as illustrated in Fig.~\ref{fig:MaxLIS lin}b. Indeed, the two most probable patterns found by Aquila shown in Fig.~\ref{fig:MaxLIS lin}c are valid MISs. Each MIS occurs with probabilities of 0.2. A subsequent 3rd most probable solution  is still an IS but not a maximum IS, as it hosts only five excitations.
%==========================================
% JAN, DO NOT ERASE IT : ~/paper_QCE2024_engineering$ ./fig4_maxLIS.py -t lin
 %========================================

\begin{figure}[htbp]
\centerline{\includegraphics[width=.99\linewidth]{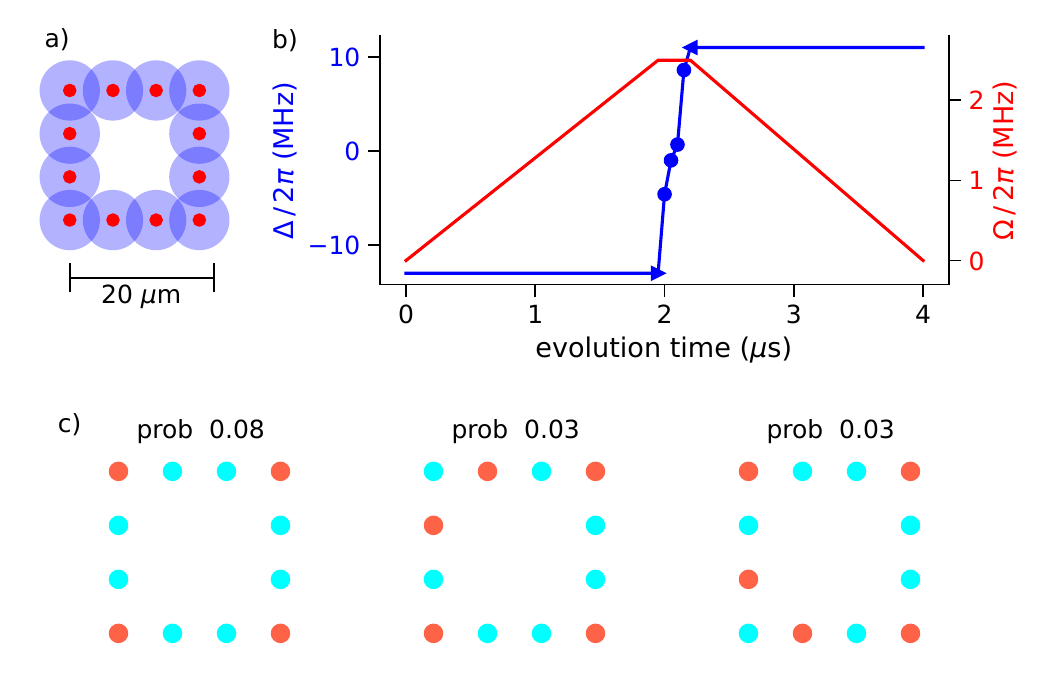}}
   \caption{MIS experiment with drive optimization. a) The placement of 12 atoms representing the corresponding graph, as explainf in Fig.~\ref{fig:MaxLIS lin}.  b) The drive optimized for promotion of the 4-atoms MaxIS. c) The experimentally obtained most probable solution is the correct one. The other solutions are at least a factor of 2.5 less probable. 
}
\label{fig:MaxLIS opt}
\end{figure}

Now, let us consider tasking Aquila with identifying a specific non-maximal MaxIS on the same 12-atom graph, specifically one characterized by the minimal number of excited atoms, one at each corner (four total). Due to the reduced cardinality and the existence of other cardinality-4 ISs, this MaxIS cannot be prepared with a sizeable probability through the adiabatic protocol used typically for the MIS problem. To achieve this new targeted pattern '100100100100', we once again employed the COBYLA optimizer to fine-tune the drives, using the probability of the target pattern as the reward function. In this optimization we introduced two additional degrees of freedom: the durations of the ramp-up and ramp-down phases were varied, while the total evolution time was maintained at $4~\mu$s. The intermediate six time steps for the detuning remained evenly distributed over the remaining time interval. We used 200,000 simulated shots per iteration. The resulting optimized drive, shown in Fig.~\ref{fig:MaxLIS opt}b, features very slow ramps of $\Omega$ and a steep rise in detuning, indicative of a diabatic transition. Remarkably, the most probable pattern Aquila identified for this diabatic transition was precisely the 4-atom MaxIS we aimed for, shown in Fig.~\ref{fig:MaxLIS opt}c. Although the probability of achieving this solution was not exceedingly high, it emerged as the most probable outcome, with a 2.5-fold higher probability than any other. 

This series of experiments underscores the adaptability of the neutral atom quantum device for addressing non-native optimization problems, particularly of the MaxIS; this is in accord with a recent comprehensive work \cite{Wurtz:2024} that applied neutral atom quantum computers to MaxCut-type graph problems.

%Further work is needed to perform a more comprehensive evaluation of optimized MaxIS %protocols in a family of graphs. 
%==========================================
% JAN, DO NOT ERASE IT : ~/paper_QCE2024_engineering$ ./fig6_maxLIS.py -t opt
 %========================================

% !TEX root = 0-main.tex
\section{Discussion}

In summary, we performed a range of quantum state engineering experiments on the neutral atom quantum processor Aquila. The single-qubit coherence time experiments showcase the reliability of the platform, as they reproduce known measurements in an independent setting \cite{wurtz2023aquila}. The two-qubit entanglement experiments performed in the weak interaction regime underscore the wide range of parameter programmability available to the quantum engineer in an analog platform. 

Our main novel results are the preparation of an antiferromagnetic state in a spin chain with one domain-wall defect and the validation of a non-native Maximal Independent Set optimization problem. Both results extensively use flexibility in programming the geometry and pulses to adapt the hardware protocols to the problem at hand. The successful preparation of the defect state might be of interest to the fundamental physics audience, as it is the first step in probing the dynamics of domain walls in a quantum phase using quench protocols \cite{Kormos:2017,Liu:2019,Tan:2021}. The non-native MaxIS validation was performed here on a small problem but potentially opens up an avenue for systematic exploration of neutral atom quantum hardware as MaxIS solvers through pulse optimization.

%%%%%%%%%%%%%%%%%%%%%%%%%%%%%%%%%%%%%%%%%%%%%%%%%%%%%%
%%%%%%%%%%%%%%%%%%%%%%%%%%%%%%%%%%%%%%%%%%%%%%%%%%%%%%

\section*{Acknowledgment}
This research was supported by and used resources of the National Energy Research Scientific Computing Center (NERSC), a U.S. Department of Energy Office of Science User Facility located at Lawrence Berkeley National Laboratory, operated under Contract No. DE-AC02-05CH11231, using NERSC award DDR-ERCAP0030190. 

\bibliographystyle{IEEEtran}
\bibliography{references}

%\appendices
%\input{8-read err corr}
\end{document}